\documentclass[journal]{IEEEtran}

\usepackage[noadjust]{cite}
\usepackage{datetime}
\usepackage{comment}
\usepackage{amsmath}
\usepackage{amssymb}
\usepackage{hyperref}
\usepackage{color}
\DeclareMathOperator*{\argmax}{argmax}

\usepackage{graphicx}
\begin{document}

\title{Taming Spontaneous Stop-and-Go Traffic Waves: \\ A Bifurcation Perspective of A Dynamical Map}
\author{Suzhou~Huang*, Jian~Hu  \\  
	\thanks{S. Huang is an independent researcher (e-mail: huang0suzhou@gmail.com). \\
		\indent J. Hu is with the University of Michigan, Dearborn, MI 48128, USA (e-mail: jianhu@umich.edu).}
	\thanks{* Corresponding Author}
}

\markboth{}%
{Taming Shockwaves}

\maketitle

\begin{abstract}
	We consider a discrete-time dynamical system in a car-following context. The system was recently introduced to parsimoniously model human driving behavior based on utility maximization. The parameters of the model were calibrated using vehicle trajectory data from the Sugiyama experiment. It was shown that such a system can accurately reproduce the observed collective phenomena of a more elaborate experiment by Tadaki et al. Once the heterogeneity and noise are switched off, the model defines a map of the corresponding discrete-time dynamical system. We first perform a bifurcation analysis of the map by studying the stability of its limit solutions: a free-flow fixed point and a stop-and-go quasi-periodic orbit. When the vehicle density is varied, our model displays a bifurcation diagram qualitatively similar to those found in a class of optimal velocity models based on an ordinary differential equation approach, including regimes where one or both of the limit solutions are stable. In a 2D bifurcation diagram we further demonstrate that imposing a vehicle density-dependent speed advisory can dissipate the stop-and-go quasi-periodic orbit. This in turn lays the mathematical foundation for a simple, yet effective proposal \cite{shen_taming_2021} to tame stop-and-go waves, improving traffic flow and smoothness simultaneously via variable speed advisory.
    
\end{abstract}

\begin{IEEEkeywords}
	human driving behavior, traffic waves, discrete-time dynamical system, bifurcation analysis
\end{IEEEkeywords}	

\IEEEpeerreviewmaketitle

\section{\bf Introduction}
Real traffic displays a variety of patterns. They range from benign, such as free-flow where vehicles are adequately spaced and moving near the imposed speed limit, to adverse, such as gridlock where vehicles grind to a complete halt. Somewhere in between there are other highly non-trivial patterns, depending on the driving environment and traffic conditions.  The case of stop-and-go waves is a good example, in which vehicles are neither in free-flow nor jam phases but alternate between the two extrema. Such waves are often induced by traffic bottlenecks.  But they can also form spontaneously, in which case they are dubbed ``phantom waves''.  It is well documented that stop-and-go waves imply multiple adverse effects to drivers (repeated acceleration and braking), vehicles (excessive wear and tear),  and the environment (higher fuel consumption and pollution).  In order to control the formation of these oscillatory waves, or to dissipate them, a mathematical model to describe the associated micro mechanism would be invaluable. To this end, a very fruitful approach is the so-called bifurcation analysis, in which a specific macro traffic pattern can be identified as a set of limit solutions for a dynamical system characterized by some ordinary differential equation (ODE).  An enormous amount of insights can be gained, at least for situations with relatively simple driving environment. Works along this line were pioneered in \cite{gasser_bifurcation_2004}, \cite{orosz_subcritical_2006}, and \cite{orosz_exciting_2009}.

Here we continue the same line of inquiry: bifurcation analysis for vehicle-following systems on a single-lane circular road. What distinguishes our work from earlier analyses are the following.  First, we deploy a human driving behavior model called {\it adaptiveSeek} recently proposed in \cite{dai_towards_2021}.  The model parameters are fixed by a systematic calibration in \cite{dai_calibration_2021} using vehicle trajectory data from the Sugiyama experiment \cite{sugiyama_traffic_2008}.  In contrast with models that specify driving policy function, such as optimal speed models \cite{bando_dynamical_1995, bando_analysis_1998}, and IDM \cite{treiber_traffic_2013}, the driving policy is derived by maximizing a judicially chosen utility function in our model.  We were able to show in \cite{dai_calibration_2021} that, in addition to fitting nicely to individual vehicle trajectories, the model can quantitatively reproduce all major collective phenomena observed in the the Tadaki experiment \cite{tadaki_phase_2013},  such as the fundamental diagram, and main characteristics of the spontaneously formed stop-and-go waves. The reason we prefer the behavior model approach, relative to the oft-deployed policy-based approach, is due to the common belief that specifying the utility function is much more parsimonious than specifying driving policy in general driving contexts beyond simple car-following.

Second,  the behavior model can be viewed as a discrete-time dynamical system, and hence defines a map, which we call the {\it adaptiveSeek} map. Once driver heterogeneity and acceleration noise are switched off, this map becomes mathematically amenable to an explicit bifurcation analysis.  We demonstrate that there are at least two types of limit solutions: a fixed point corresponding to free-flow traffic, and a quasi-periodic orbit corresponding to stop-and-go waves. We then examine the stability of these limit solutions with a combination of linearization and simulation.  Because we are dealing with a discrete-time map with delay rather than ODEs, we need to invoke somewhat different mathematical techniques than those typically found in literature.  Amazingly, despite seemingly different micro model specifications, we find very similar results in \cite{gasser_bifurcation_2004}, \cite{orosz_subcritical_2006},  and \cite{orosz_exciting_2009}. These include 1) a bifurcation diagram as vehicle density ($\rho$) is varied; 2) Neimark-Sacker bifurcation (discrete-time analog of Hopf bifurcation) that breaks the linear stability of the map; and 3) the existence of bi-stable regimes where both free-flow and stop-and-go solutions are asymptotically stable.

Third, we have gone one step further than stability analysis by extending the usual one-dimensional bifurcation analysis in vehicle density to two-dimensional bifurcation analysis on both vehicle density ($\rho$) and another parameter of the model: the ideal speed ($v^*$) of each vehicle. The latter parameter can be controlled from a traffic management perspective by imposing a $\rho$-dependent variable speed advisory (VSA). We use simulations to partition the ($\rho,\,v^*$)-plane into regimes depending on which of the limit solutions is asymptotically stable. We further demonstrate that the optimal way to tame spontaneously formed stop-and-go waves is practically equivalent to locating a 2D bifurcation curve where the only stable limit solution is the free-flow fixed point.  In so doing we lay the mathematical foundation for the same task achieved using the computational mechanism design approach in \cite{shen_taming_2021}.

The stability analysis of traffic dynamics, dating back to the pioneering car-following models from the 1950s and early 1960s \cite{pipes_operational_1953, chandler_traffic_1958, herman_traffic_1959, gazis_nonlinear_1961}, interprets the propagation of disturbances down a line of vehicles. Those microscopic car-following models describe the time evolution of individual vehicles using ordinary differential equations (ODEs) and, on this basis, delineate the dynamical state of the system. Some modifications related to classical car-following models have been investigated in \cite{unwin_stability_1967, zhang_stability_1997} and earlier works referenced therein. Bando et al. \cite{bando_dynamical_1995, bando_analysis_1998} propose the optimal velocity (OV) model and further take into account the reaction-time delay of drivers. They investigate the linear stability of steady-state traffic flow on a circular road in the context of the OV models and demonstrate a transition from free flow to traffic congestion as an effect of the increase in the average vehicle density. The same phenomenon is verified in both the Sugiyama and Tadaki experiments \cite{sugiyama_traffic_2008, tadaki_phase_2013}. Gasser et al \cite{gasser_bifurcation_2004} and Orosz et al. \cite{orosz_subcritical_2006, orosz_exciting_2009} theorize about the stability of the OV models using bifurcation theory. Gasser et al \cite{gasser_bifurcation_2004} ascribe the loss of stability to a Hopf bifurcation and discover the coexistence of the linear stability of a fixed point and a stable periodic orbit, i.e. stop-and-go wave. Orosz and St\'{e}p\'{a}n \cite{orosz_subcritical_2006} perform an analytical Hopf bifurcation calculation with regard to the average vehicle density in the presence of drivers’ reaction delay. Moreover, Orosz et al. \cite{orosz_exciting_2009} also reveal the bi-stable phenomenon that, while the steady-state flow is linearly stable, stop-and-go traffic jams can be triggered by sufficiently large localized perturbations. They interpret this phenomenon as as something caused by the existence of unstable small-amplitude oscillations separating the free flow and the large-amplitude stop-and-go waves. 

In the past decade, studies extend to the stability analysis of various car-following models. Li and Ouyang \cite{li_characterization_2011} develop a mathematical framework to quantify the period and magnitude of traffic oscillations for a general class of nonlinear car-following laws and analyze these characteristics across the propagation of traffic oscillations. Yu et al. \cite{yu_new_2014} analyze  the stability of a model with time delays in sensing headway and velocity. Jin and Xu \cite{jin_stability_2016} add the delayed feedback control in the OV models with reaction-time delay and state the stability conditions of this modification. Tomoeda et al. \cite{tomoeda_bifurcation_2018} derive linear stability conditions depending on the relative velocity parameterized in the STNN model proposed in \cite{shamoto_car-following_2011}. Zhang et al. \cite{zhang_bifurcation_2019} discuss the stability of a newly developed car-following model considering the velocity difference in a time interval. Ngoduy and Li \cite{ngoduy_hopf_2021} establish a general bifurcation structure of a generic car-following model, which embodies models of OV, full velocity difference, intelligent driver, and cooperative adaptive cruise control. They discuss the impact of the multiple time delays in sensing distance headway, speed, as well as relative speed. Petersik et al. \cite{petersik_one-parametric_2021} use an equation-free method to perform bifurcation analyses of artificial neural network-based car-following models. The aforementioned studies mainly use Fourier or Laplace transform methods to derive stability conditions for continuous-time car-following models. Qin et al. \cite{qin_scalable_2017, qin_stability_2017} apply the stability analysis into connected vehicle systems. They discretize the time interval and the string stability is analyzed on the $z$-domain.  In addition, other studies investigate how stop-and-go waves can be dissipated by adding automated vehicles to the ring. Chou et al. \cite{chou_lord_2022} review well-known autonomous control models --- such as the augmented OV-FTL model \cite{cui_stabilizing_2017}, the bilateral control model \cite{horn_suppressing_2013}, the FollowerStopper controller \cite{stern_dissipation_2018}, the proportional-integral with saturation controller \cite{delle_monache_feedback_2019, stern_dissipation_2018}, Optimal Control Strategy \cite{zheng_smoothing_2020}, Lyapunov-based controller \cite{delle_monache_feedback_2019}, reinforce learning control \cite{wu_framework_2017, wu_tracking_2019}, etc. --- and carry out the simulation experiments to evaluate their performance.

The remainder of our paper is organized as follows. In the next section we first briefly introduce the human driving behavior model as a map of a discrete-time dynamical system. In Section \ref{limit_solutions}, we discuss and illustrate the qualitative nature of the limit solutions of the dynamical system, which are a free-flow fixed point and a stop-and-go quasi-periodic orbit. We then investigate the linearized version of the dynamical system and study its stability at the free-flow fixed point using the $z$-transform technique in Section \ref{linear_stability}. We show that the dynamical system loses its linear stability via Neimark-Sacker bifurcations when the vehicle density is varied. Then we elucidate the bifurcation behavior of the stop-and-go quasi-periodic orbit via simulation study and sketch out the corresponding bifurcation diagram in Section \ref{bistable_regime}. The computed bifurcation diagram clearly shows that there are bi-stability regimes where both limit solutions can exist, depending on the initial condition. In Section \ref{2D_bifurcation} we show a practical method for taming stop-and-go traffic waves by imposing a vehicle density-dependent speed advisory.  Mathematically, this can be accomplished with a two-dimensional bifurcation analysis on the $(\rho,v^*)$-plane. Here we demonstrate that when the ideal speed $v^*$ is below a threshold, the stop-and-go quasi-periodic orbit loses its stability. This in turn implies an effective strategy for dissipating traffic waves in a Pareto-optimal manner. A summary and outline for future work are presented in the last section.
 
\section{\bf The Human Driving Behavior Model and Map}\label{BehavioralModeling}
In \cite{dai_towards_2021} we proposed a heuristics-based driving decision algorithm called {\it adaptiveSeek}. In this approach a Markov game with continuous actions and simultaneous moves was used as the modeling framework. The solution concept {\it adaptiveSeek} was simplified from a rigorous game theory-based approach dubbed {\it betaNash} by relaxing some of the theoretical requirements that are too restrictive, both conceptually and computationally. It was shown explicitly that {\it adaptiveSeek} can well aproximate the sub-game perfect Nash equilibrium solutions while taking into account the bounded rationality, as illustrated in a situation of mandatory lane changes on a double-lane highway.  It was also shown in \cite{dai_calibration_2021} that the model parameters for traffic on a circular road in the Sugiyama setting can be quantitatively calibrated using observed vehicle trajectory data.  Furthermore, it was explicitly demonstrated that the calibrated model can be generalized to the setting of the Tadaki experiment \cite{tadaki_phase_2013}, where vehicle density is varied, in the sense that it can reproduce nearly all the observed collective phenomena well \cite{shen_taming_2021}. Therefore, we will regard {\it adaptiveSeek} as the decision-making model for human driving behaviors. For mathematical tractability of our bifurcation analysis we switch off all the driver heterogeneity and state evolution noise.  As shown in \cite{shen_taming_2021} these details do not affect the qualitative results, so long they are not too big. 

\subsection{\bf The behavioral model specification}
Here we recapitulate the decision-making algorithm {\it adaptiveSeek} introduced in \cite{dai_towards_2021,dai_calibration_2021}. Specific considerations and justifications can be found in the original papers. Only longitudinal dynamics are explicitly modeled.
\subsubsection{The traffic setting}
There are $N$ vehicles in the system traversing on a circular road of circumference $C$. The vehicle ordering is chosen so that vehicle $i$ is always behind vehicle $i+1$. Periodic boundary condition is imposed, i.e. all position variables $x_{i,t}$ (and headway) should be understood as $\pmod{C}$, and the $(N+1)$-th vehicle is identified as the first vehicle.  For simplicity, we use {\it agent} to refer either vehicle, driver,  or a combination of both.
\subsubsection{Driving decision-making}
There are two kinds of state variables that are conceptually distinct. The first kind is for characterizing the kinematics of vehicles, and the second for drivers' decision-making. The kinematic state for vehicle $i$ is defined by a vector $\xi_{i,t}=(x_{i,t},v_{i,t},a_{i,t})^\top$,  whose components represent the position, velocity and acceleration of the vehicle at time $t$, respectively.  The decision-making state by the driver of vehicle $i$ at time $t$ is defined by $s_{i,t}=(\xi_{i,t}^\top,\xi_{j,t}^\top)^\top$\footnote{Conceptually, $\xi$'s components appearing in $s_{i,t}$ could be different from the $\xi$'s for characterizing vehicles' motion, due to drivers' estimation error. Since bifurcation analysis precludes us from including driver heterogeneity and noise, we will ignore such differences in this work.}, involving the ego vehicle $i$ and the leading vehicle $j\equiv i+1$. 

The kinematic state evolution can be written as
\begin{equation}
	\begin{cases}
		x_{i,t+1}&= x_{i,t}+v_{i,t}\,\Delta t \,\,\,\pmod{C}\\
		v_{i,t+1}&=v_{i,t}+a_{i,t}\,\Delta t\\
		a_{i,t+1} &=\gamma\,a_{i,t}+ \big({u}_{i,t}- \gamma\,{u}_{i,t-1} \big)
	\end{cases}\, ,
	\label{VehicleDynamics}
\end{equation}
with $u_{i,t}$ being the control input (or action) of the driver. Note that the acceleration is modeled as an AR(1) process in order to capture its stickiness in vehicle dynamics.

Decision-making is driven by utility maximization. In a game setting, the utility is not only a function of its own control input $u_{i,t}$ but is also dependent on control inputs from all other interacting neighboring agents $u_{-i,t}$. In order to relax the strong assumptions associated with dynamic Nash equilibria so as to more closely mimic bounded rationality of human driving, additional behavioral assumptions are made appropriate for car-following: constant action for the ego agent ($u_{i,t}=u$); zero action for other agents ($u_{-i,t}=0$). Thus, the driving policy for driver $i$ can be expressed as
\begin{equation}
	u^*_{i,t}(s_{i,t})=\argmax_{u\in[u_\text{min},u_\text{max}]} 
	U^\text{eff}_{i,t}(u|s_{i,t})\, . \label{BestResponseTilde}
\end{equation}
The effective utility typically consists of several utility components, each of which takes care of a specific aspect of driving preference. It is
further related by multiple per-period utilities evaluated at anticipated future states within the planning horizon $H+1$: $\forall h\in\{0,\cdots,H\}$
\begin{equation}
	U^\text{eff}_{i,t}(u|s_{i,t})=\sum_{k}\, w_{i,k} \,g_k\Big[\cup_{h=0}^H U_{i,t}^{(k)}(u|\hat{s}_{i,h})\Big]_{\hat{s}_{i,0}=s_{i,t}}\, ,
	\label{EffectiveUtility}
\end{equation}
where $U_{i,t}^{(k)}$ is the $k$-th component of the per-period utility with weight $w_{i,k}$. Instead of the standard cumulative utility form, we use transformed utility with $g_k[...]$. This transformation is necessary for better fitting human driving trajectories. Starting from the current state $\hat{s}_{i,0}=s_{i,t}$, the anticipated future state for the ego agent $i$ evolves according to
\begin{equation}
	\begin{cases}
		\hat{x}_{i,h+1}&=\hat{x}_{i,h}+\hat{v}_{i,h}\,\Delta t \,\,\,\pmod{C}\\
		\hat{v}_{i,h+1}&=\hat{v}_{i,h}+\hat{a}_{i,h}\,\Delta t\\
		\hat{a}_{i,h+1}&=u
	\end{cases}\, ,
	\label{anticipation_evolution}
\end{equation}
and similarly for other interacting agents ($-i$) with zero action.

Conceptually, it is important to distinguish the two state evolutions defined in Eq.(\ref{VehicleDynamics}) and Eq.(\ref{anticipation_evolution}). The former models the mechanical realization for vehicle $i$ given driver $i$'s control inputs, whereas the latter models driver $i$'s mental anticipation of future states for all relevant agents. 

\vskip 0.1in
\subsubsection{Utility function and its parameterization}
For our car-following setting we only need three utility components. In the following, we utilize the available information to the fullest, in the sense that by knowing $\hat{s}_{i,h}=(\hat{x}_{i,h},\hat{v}_{i,h},\hat{a}_{i,h})$ we also know $\hat{x}_{i,h+1}=\hat{x}_{i,h}+\hat{v}_{i,h}\Delta t$ and $\hat{v}_{i,h+1}=\hat{v}_{i,h}+\hat{a}_{i,h}\Delta t$.
\subsubsection*{Moving forward reward at a desired speed}\label{Moving_forward}
The first component represents the intention to move forward along the circular road at the desired speed:
\begin{equation}
	U_{i, t}^{(1)}(u|\hat{s}_{i,h})=\exp\Big(-\Big(\frac{\hat{v}_{i, h+1}+u\,\Delta t- v_i^*}{\kappa_i^{(1)} v_i^*}\Big)^{2}\Big)\, ,
\end{equation}
where $v_i^*$ is the ideal speed and $\kappa_i^{(1)}$ controls the degree by which agent $i$ likes to be close to its ideal speed. 

The anticipation related transformation function for this first component is given by 
$$g_1\big[\cup_{h=0}^H\,U_{i, t}^{(1)}(u\big|\hat{s}_{i,h})\big]=U_{i, t}^{(1)}(u|\hat{s}_{i,h})\big|_{h=0} \, ,$$
ie. from all $H+1$ items sequentially only the first matters.

\subsubsection*{Moving backward penalty}\label{Moving_backward}
The second component is a penalty for moving backward:
\begin{equation}
 U_{i, t}^{(2)}(u|\hat{s}_{i,h})=\exp\Big(-\kappa_v^{(2)}\big(\hat{v}_{i, h+1}+u\,\Delta t +\kappa_0^{(2)}\big)\Big)\, .
\end{equation}
This term is needed in order to prevent vehicle velocity to be persistently negative. Also, $g_2$ is chosen similarly as $g_1$:
$$g_2\big[\cup_{h=0}^H\,U_{i, t}^{(2)}(u\big|\hat{s}_{i,h})\big]=U_{i, t}^{(2)}(u|\hat{s}_{i,h})\big|_{h=0} \, .$$

\subsubsection*{Pairwise collision penalty}\label{Pairwise_collision}
The third component represents the subjective risk perceived by the driver for one-on-one collisions with another vehicle. Let $L_i$ denote the length of vehicle $i$, and let $\mathcal{F}(x)=\exp{( -x^2-2x )}$. For vehicle $i$'s front collision, we choose
\begin{equation}
	U_{i, t}^{(3)}(u|\hat{s}_{i,h})=\left\{
	\begin{array}{ll}
		1, & \Delta x_{i, j, h} \leq 0 \\ [6pt]
		\mathcal{F} \Big(\displaystyle\frac{\Delta x_{i, j, h}}{\delta_{i,j, h}}\Big), & 0<\Delta x_{i, j, h}
	\end{array}\right. 
	\label{PairwiseCollisionPenalty}
\end{equation}
where $\Delta x_{i,j,h}=(\hat{x}_{j,h+1}+\hat{v}_{j,h+1}\Delta t-L_{j}/2)-(\hat{x}_{i,h+1}+\hat{v}_{i,h+1}\Delta t+L_i/2)$ is the bumper-to-bumper distance of vehicle $i$ to its leading vehicle $j$. It is natural to choose the front scale parameter to be speed dependent: $\delta_{i,j,h} = \kappa_{i,c}^{(3)}+\kappa_{i,v}^{(3)}|\hat{v}_{i,h+1}+u\Delta t | + \kappa_{i,d}^{(3)}\max\{\hat{v}_{i,h+1}+u\Delta t-\hat{v}_{j,h+1},0\}$. The $\kappa_{i,d}^{(3)}$ term is needed for braking when the vehicle ahead is slower. Because of the discouragement for moving backward in $U_{i, t}^{(2)}$ we do not need to explicitly consider the rear collision penalty here.
The anticipation-related transform function for the third component is given by 
$$g_3\big[\cup_{h=0}^H\,U_{i, t}^{(3)}(u\big|\hat{s}_{i,h})\big]=\max_{h\in\{0,\cdots,H\}} U_{i,h}^{(3)}\big(u|\hat{s}_{i,h}\big)\, .$$

\vskip 0.1in
\subsubsection{Model parameters}
Our goal is to have a realistic driving behavior model for the Tadaki setting,  though with agent heterogeneity and state evolution noise switched off.  Ideally, we should have calibrated model parameters if we had vehicle trajectory data from the Tadaki experiment.  Since this is not the case, the closest data we can find is that of the Sugiyama experiment. Therefore, we adopted the same behavior model and its model parameters derived from the Sugiyama setting \cite{dai_calibration_2021}.  As is done in \cite{shen_taming_2021},  we then average over all 22 Sugiyama agents to arrive the following common model parameters for this study: $v^*=10.49$ (m/s),  $\kappa^{(1)}=0.7$,  $w^{(1)} =1$; $\kappa_v^{(2)}=10$,  $\kappa_0^{(2)}=0.25$ (m/s),  $w^{(2)}=-1$; $\kappa_c^{(3)}=0.6$ (m),  $\kappa_v^{(3)}=0.3$ (s),  $\kappa_d^{(3)}=1.0$ (s), and $w^{(3)}=-10$.  We further choose $\Delta T=1/6$ (s), $\gamma=\sqrt{0.7}$, and $H=7$ in our specific context, which leads to a look-ahead time horizon of $(H+1)\,\Delta t=4/3$ (s).  All vehicles have the same length: $L=3.9$ (m).  Control input is limited as $u\in\mathcal{U}\equiv[u_\text{min},u_\text{max}]=[-6,4]$ (m/s$^2$).

\subsection{\bf The map of {\it adaptiveSeek}} 
For mathematical convenience we approximate \eqref{BestResponseTilde} via the following Boltzmann weighted average:
\begin{equation}
	\bar{u}_{i,t}(s_t)\equiv \sum_{u} u P_{i,t}(u|s_{i,t})\, ,
	\label{u_bar}
\end{equation}
with $P_{i,t}(u|s_{i,t})\propto\exp[\lambda\, U^\text{eff}_{i,t}(u|s_{i,t};h)]$.  When $\lambda$ is positive and large, $\bar{u}_{i,t}(s_t)\rightarrow u^*_{i,t}(s_{i,t})$.  Thus, we can regard the control input in Eq.(\ref{u_bar}) as a regularized version of Eq.(\ref{BestResponseTilde}),  such that $\bar{u}$ has a continuous support and is differentiable with respect to model parameters, even when the utility maximization is done via grid search.\footnote{If one insists on using the usual utility maximization, which is computationally more involved than a much easier grid search, the partial derivatives of $u^*(s_{i,t})$ with respect to the state variable can still be calculated using implicit differentiation on the first order condition of utility maximization. This argument can be formalized by Laplace approximation when $\lambda\to\infty$. The grid we choose has 41 points for $u\in[-6,4]$ with $\lambda=200$.}

The endogenized state evolution in Eq.\eqref{VehicleDynamics} for vehicle $i$ can then be expressed as follows:
\begin{equation}
	\begin{cases}
		x_{i,t+1}&= x_{i,t}+v_{i,t}\,\Delta t \,\,\,\pmod{C}\, ,\\
		v_{i,t+1}&=v_{i,t}+a_{i,t}\,\Delta t\, , \\
		a_{i,t+1} &=\gamma\,a_{i,t}+ \big(\bar{u}_{i,t}(s_{i,t})- \gamma\,\bar{u}_{i,t-1}(s_{i,t-1}) \big)\, .
	\end{cases}\label{map_evolution}
\end{equation}
The above state evolution is executed by all agents in parallel step-by-step, as implied by the Markov game setting. In so doing, every agent is treated symmetrically, in the sense that each vehicle is simultaneously an ego vehicle in its own state evolution, but also can appear as a surrounding vehicle in other agents' utility calculation.  

Putting the state for every vehicle together we obtain a discrete time map in the 3N-dimensional collective state space: $\boldsymbol{\xi}_{t}\equiv\big(\xi_{1,t}^\top,\cdots,\xi_{N,t}^\top\big)^\top$,
\begin{equation}
	\boldsymbol{\xi}_{t+1}=\boldsymbol{f}\big(\boldsymbol{\xi}_{t},\,\boldsymbol{\xi}_{t-1}\big). 
	\label{map_adaptiveSeek}
\end{equation}
Because the delay only appears in acceleration, not in position and velocity, our map is really $\boldsymbol{f}: \mathbb{R}^{4N}\mapsto\mathbb{R}^{4N}$.\footnote{The dimensionality of the map can be seen more clearly by directly augmenting the state space: $b_{i,t+1}=\bar{u}_{i,t}(s_t)$ as an auxiliary state variable for each agent.  However, we will keep $\boldsymbol{\xi}_{i,t}$ to only represent the physical state for convenience in visualization.} Of course, the map also depends on all model parameters implicitly. 

We are particularly interested in finding how the limit behaviors of the map change qualitatively when some of the externally influenceable parameters are varied, such as the vehicle density $\rho$ and ideal speed $v^*$. Note that 1) the control input $\bar{u}$ is the only nonlinearity of the map, and 2) there is a one period of lag in Eq.(\ref{map_evolution}) and Eq.(\ref{map_adaptiveSeek}). We want to emphasize that this very short time lag in itself should not be interpreted as the explicit delay of human reaction time with respect to the observed state.  Instead, it is the induced AR(1) process, together with the anticipation process, that actually models the human driving behaviors dynamically.  We note that the explicit lag requires us to deploy a more specialized methodology ($z$-transform for general difference equations) than the standard eigen-analysis when we study the linear stability.

\section{\bf Qualitative nature of limit solutions}\label{limit_solutions}
We have detected two types of qualitatively different solutions that are asymptotically stable by simulating the map defined in 
Eq.(\ref{map_adaptiveSeek}). Generally, when both the vehicle density and ideal speed are low, the solution tends to converge to a free-flow fixed point, in which all vehicles are equally spaced with constant velocity.  When the vehicle density and ideal speed are high, the solution tends to converge to a stop-and-go quasi-periodic orbit. For illustration purposes we run simulations with the following parameter values in this section: $N=28$,  and $C=314$ (m),  while only varying the ideal speed $v^*$. The initial condition is such that all vehicles are equally distanced moving with a constant velocity equal $v^*-1$. The system is autonomous after we apply an exogenous kick to the first vehicle (constant braking with $u=-1$ (m/s$^2$) for 6 seconds while its velocity is positive) so as to provide a seed for the possible spontaneous formation of stop-and-go waves in the long-run limit. 

\begin{figure}[!h]
	\centering
	\begin{tabular}{lr}
		\includegraphics[width=.44\textwidth]{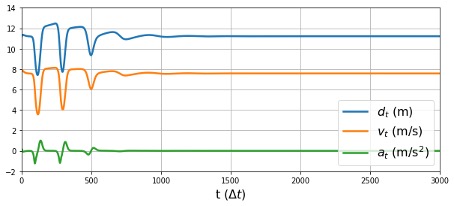}
	\end{tabular}
	\caption{A free-flow solution (with $v^*=9.0$ (m/s)) for vehicle \#14: time series of distance (center-to-center) to the vehicle ahead ($d_t$, blue), velocity ($v_t$, orange), and acceleration ($a_t$, green).  The initial transient due to the kick dies out gradually, and a steady limit is approached asymptotically.  Time is in units of $\Delta t=1/6\;(\text{s})$.}
	\label{timeSeries_FF}
\end{figure}

\begin{figure}[!h]
	\centering
	\begin{tabular}{lr}
		\includegraphics[width=.47\textwidth]{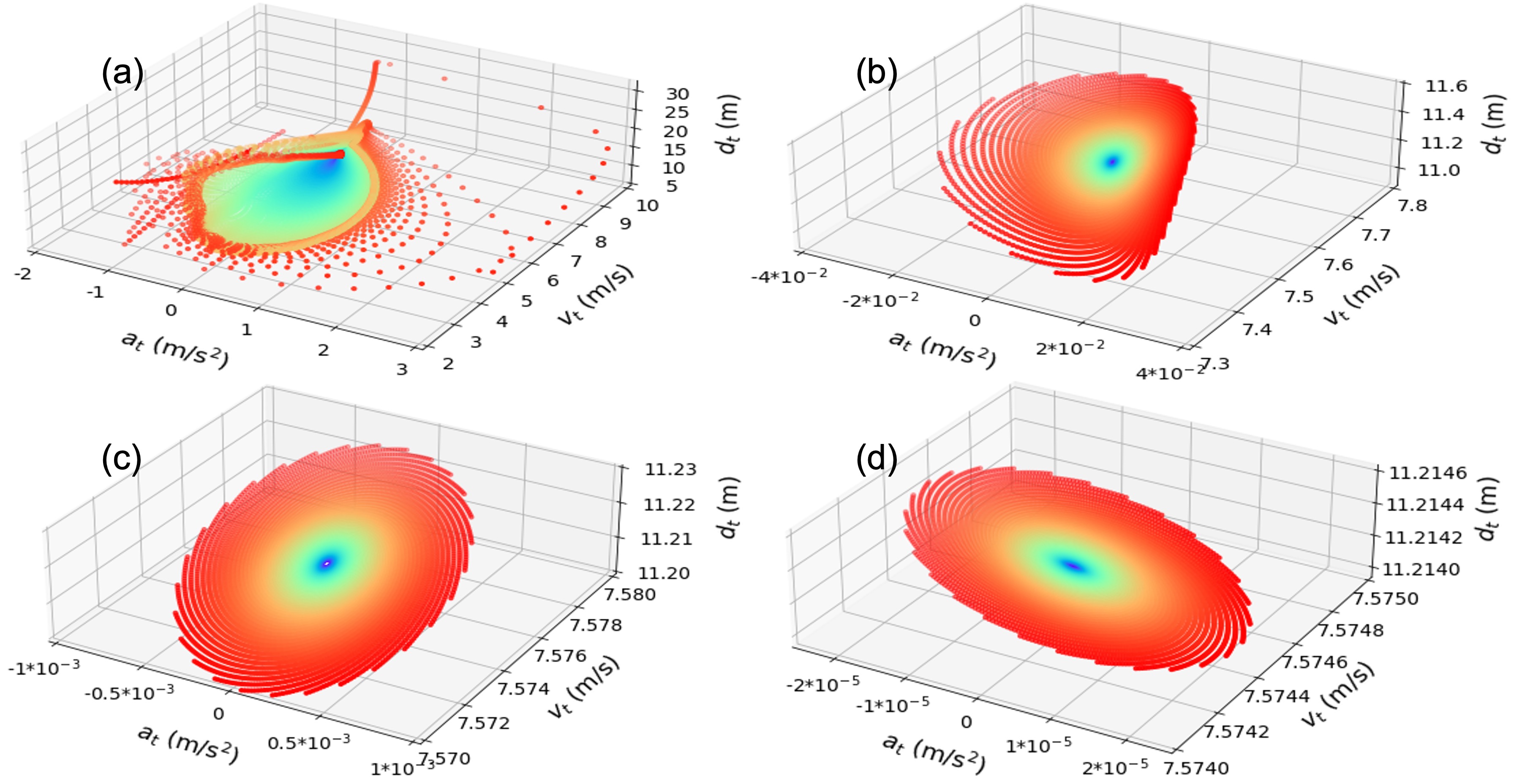}
	\end{tabular}
	\caption{A free-flow solution (with $v^*=9.0$ (m/s)) for all vehicles: phase diagrams (a) $t\in[1,750]$, (b) $t\in[751,1500]$, (c) $t\in[1501,2250]$, (d) $t\in[2251,3000]$.  All trajectories converge towards a fixed point.   Time flows from warm colors to cold colors in each panel. }
	\label{limitSet_FF}
\end{figure}

A free-flow solution is shown in Fig.\ref{timeSeries_FF} with $v^*=9.0$ (m/s) for vehicle \#14 (all other vehicles behave similarly but time-shifted). Despite the initial kick at the beginning of the simulation, the system gradually settles to a steady limit. To confirm that the steady limit is indeed approaching a fixed point, we show the trajectories for all the vehicles in the state space in Fig.\ref{limitSet_FF}. The convergence process appears to be modulated by some oscillatory modes spiraling exponentially towards the fixed point. 

\begin{figure}[!h]
	\centering
	\begin{tabular}{lr}
		\includegraphics[width=.44\textwidth]{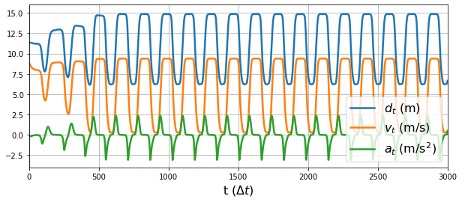}
	\end{tabular}
	\caption{A stop-and-go wave solution (with $v^*=10.0$ (m/s)) for vehicle \#14: time series of distance (center-to-center) to the vehicle ahead ($d_t$, blue), velocity ($v_t$, orange), and acceleration ($a_t$, green). The kick induces a persistent stop-and-go wave that appears to be ``periodic'' from casual observation. Time is in units of $\Delta t=1/6\;(\text{s})$. }
	\label{timeSeries_SG}
\end{figure}

\begin{figure}[!h]
	\centering
	\begin{tabular}{lr}
		\includegraphics[width=.48\textwidth]{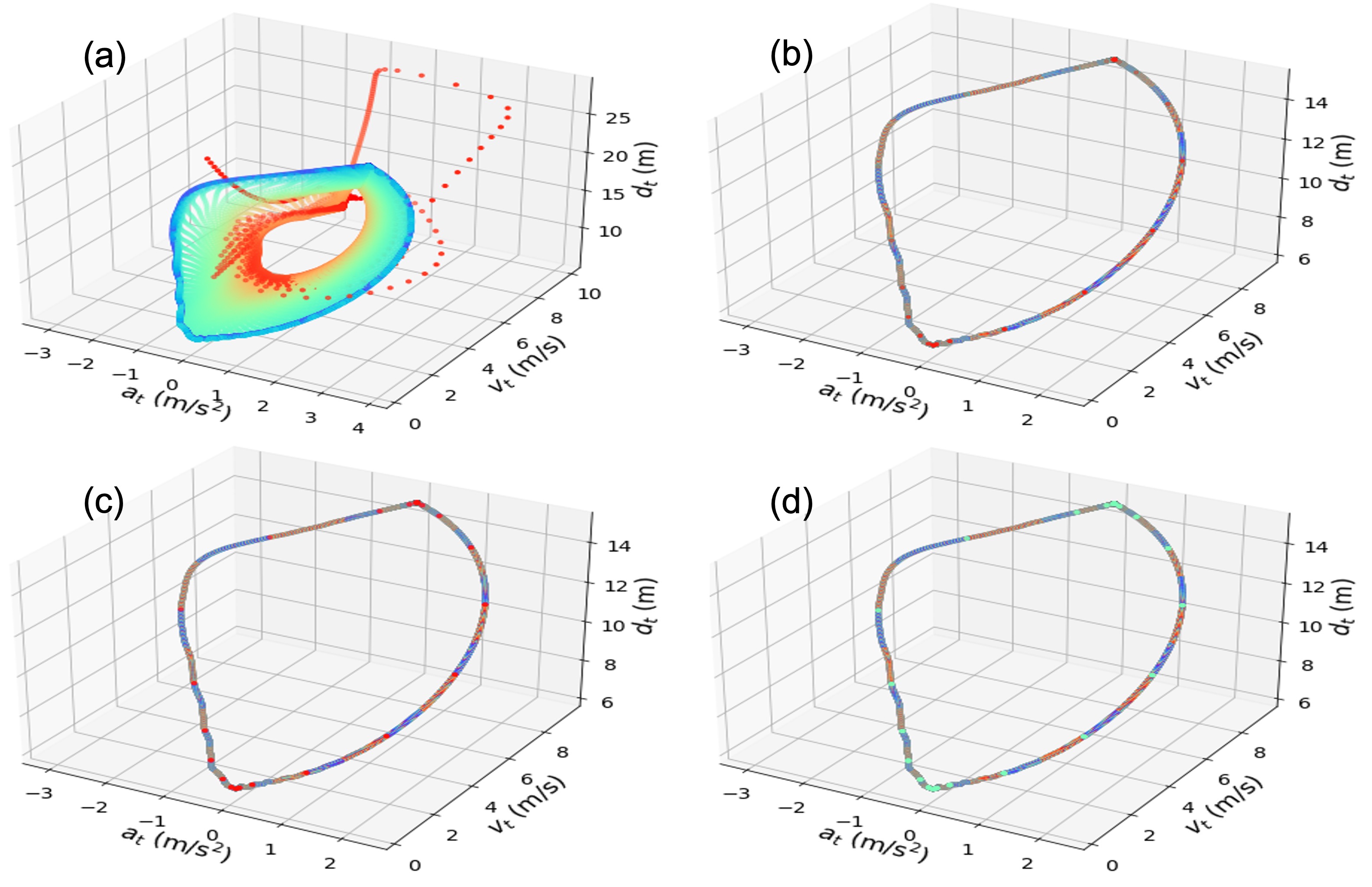}
	\end{tabular}
	\caption{A stop-and-go solution (with $v^*=10.0$ (m/s)) for all vehicles: phase diagrams (a) $t\in[1,750]$, (b) $t\in[751,1500]$, (c) $t\in[1501,2250]$, (d) $t\in[2251,3000]$. All trajectories converge to an invariant curve.  Time flows from warm colors to cold colors in each panel.  Note that the color patterns in (b), (c), and (d) are irregular, even though the invariant curve is the same.}
	\label{limitSet_SG}
\end{figure}

A stop-and-go wave solution is shown in Fig.\ref{timeSeries_SG} with $v^*=10.0$ (m/s), again for vehicle \#14. The initial kick is able to induce an oscillatory wave initially and then the system gradually settles into a persistent ``periodic'' pattern for all practical purposes. To elucidate the nature of the limit behavior in this case, we show the trajectories for all the vehicles in the state space in Fig.\ref{limitSet_SG}. The long-run limit does not look like a true periodic solution, but more like a quasi-periodic orbit on an invariant curve, signaled by the sporadic color pattern of the trajectory points (i.e. lack of systematic color pattern) on the limit set.  Since we have imposed the periodic condition in $x_{i,t}$, one possible explanation for the quasi-periodic nature of the stop-and-go wave is that the specific dynamics are not commensurate (or mode-locked, see \cite{sarmah_neimark-sacker_2014} for an example of a quasi-periodic limit solution in the delayed logistic map) with the ring road circumference, analogous to orbits on a 2D invariant torus with irrational periodicity ratio in each direction. 

It is important to point out that these two types of asymptotically stable solutions are qualitatively robust against heterogeneity among agents and noise in state evolution, since very similar patterns were observed in more realistic simulations from \cite{shen_taming_2021},  so long as the magnitudes of heterogeneity and noise are not too large.  We have seen in previous works, such as the delayed logistic map and the Henon map, that even in simple two-dimensional maps, intricate behaviors arise from their one-dimensional delayed counterparts. Thus, it is no surprise that with our own more complex map, we find similar gross features. Despite the ``sporadic'' behavior locally, the quasi-periodic orbit in Fig.\ref{limitSet_SG} is globally stable in the sense that points stay on the invariant curve once they have reached it. This is consistent with the seemingly ``periodic'' solution we observe in Fig.\ref{timeSeries_SG}.

\section{\bf Linear stability: {\it perturbation study}}\label{linear_stability}
The linear stability of a map near a fixed point is determined by the eigen-spectrum property of its Jacobian evaluated at the fixed point \cite{kuznetsov_elements_2004}. However, due to the delay introduced in the AR(1) process, we find it convenient to invoke an alternative approach: $z$-transform.  In this section we concentrate on the vehicle density dependence without the variable speed advisory (i.e.  $v^*=10.49$ (m/s)).

\subsection{\bf The free-flow fixed point}
The fixed point associated with free flow, i.e. equally spaced constant motion for all agents, can be parameterized as follows:
$ \forall i\in\{1,\cdots,N\}$,
\begin{equation} 
	\begin{cases}
		x_{i,t}^*&=x_{i+1,t}^* -1/\rho\,\,\,\pmod{C}\, ,\\
		v_{i,t}^*&=v_{0}\, , \\
		a_{i,t}^* &=0\, ,
	\end{cases}\label{FF_fp}
\end{equation}
where $\rho\equiv N/C$ is the vehicle density. The common equilibrium velocity $v_0$ is derived from  
\begin{equation}
	\bar{u}_{i,t}(s^*_{i,t})=\text{softmax}_{u\in\mathcal{U}}\,U^\text{eff}_{i,t}(u|s^*_{i,t};h)=0\, ,
	\label{equilibrium_velocity}
\end{equation}
with the equilibrium state defined as 
$s^*_{i,t}= ({\xi^*_{i,t}}^\top,{\xi^*_{i+1,t}}^\top)= (x^*_{i,t},v_0,0,x^*_{i,t}+1/\rho,v_0,0)$.  Since the effective utility function $U^\text{eff}_{i,t}$ depends only on the relative distance between two successive vehicles instead of their absolute positions, Eq.(\ref{equilibrium_velocity}) implies that the equilibrium velocity is a function of vehicle density $\rho$ and ideal speed $v^*$ only, i.e. $v_0=v_0(\rho,v^*)$, given all other fixed model parameters.

\begin{figure}[!h]
	\centering
	\begin{tabular}{lr}
		\includegraphics[width=.44\textwidth]{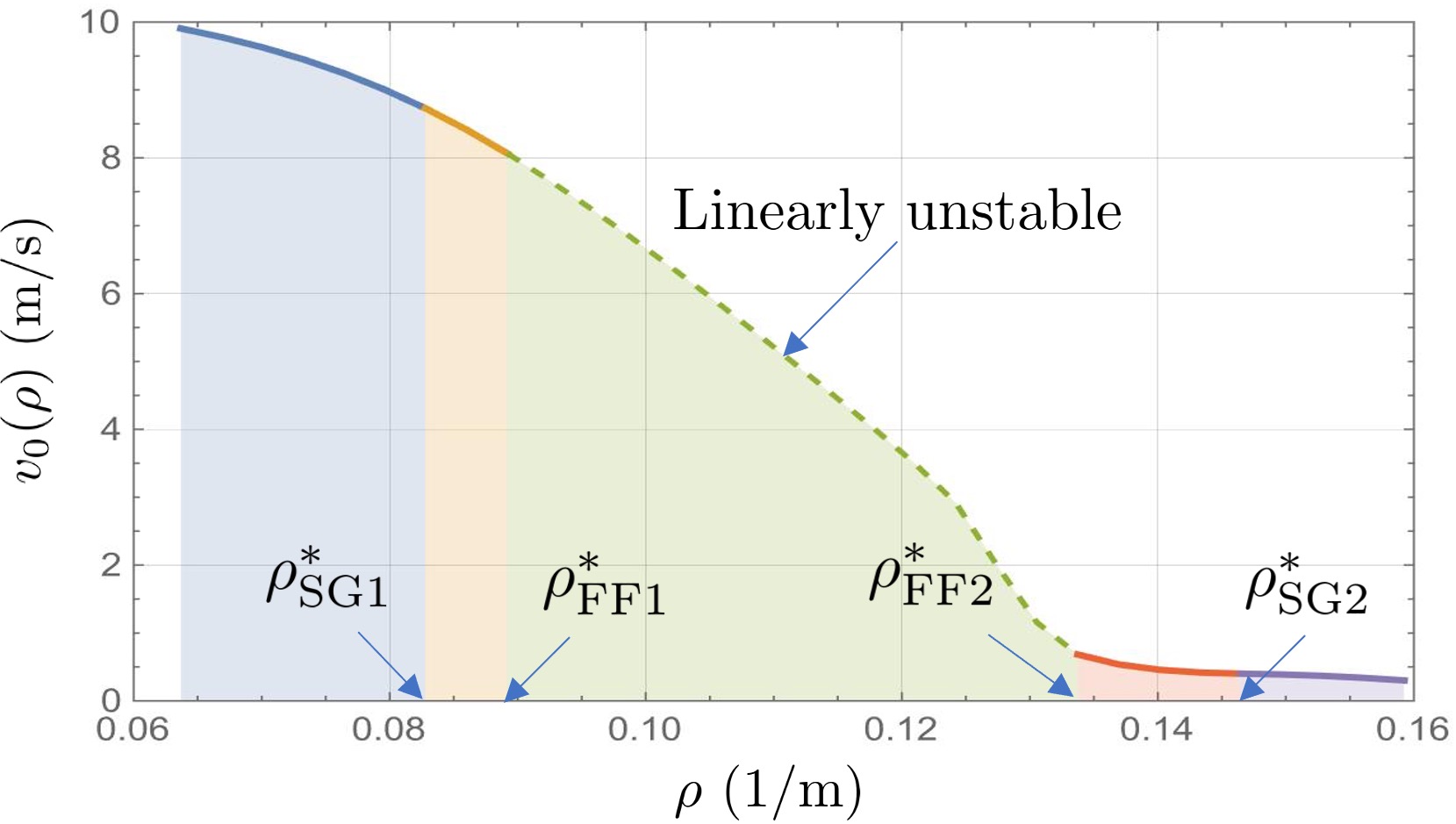}
	\end{tabular}
	\caption{The equilibrium velocity $v_0(\rho)$ as a function of the vehicle density $\rho$ at the free-flow fixed point. The free-flow fixed point is linearly stable when $\rho<\rho^*_\text{FF1}$ or $\rho>\rho^*_\text{FF2}$. The stop-and-go wave is stable when $\rho\in(\rho^*_\text{SG1},\rho^*_\text{SG2})$. There are two bi-stable regimes: $\rho\in(\rho^*_\text{SG1},\rho^*_\text{FF1})$ and $\rho\in(\rho^*_\text{FF2},\rho^*_\text{SG2})$ where both limit solutions are stable.}
	\label{FF_v0_plot}
\end{figure}

In Fig.\ref{FF_v0_plot} we show the numerical result for the equilibrium velocity as a function of the vehicle density: $v_0(\rho)$ at the free-flow fixed point.  As we will find out later in this section and the next two, there are five regimes in terms of vehicle density, separated by four bifurcation points: $\rho^*_\text{SG1}<\rho^*_\text{FF1}<\rho^*_\text{FF2}<\rho^*_\text{SG2}$. The free-flow fixed point is linearly stable when $\rho<\rho^*_\text{FF1}$ or $\rho>\rho^*_\text{FF2}$. The stop-and-go wave is stable when $\rho\in(\rho^*_\text{SG1},\rho^*_\text{SG2})$. There are two bi-stable regimes where both the free-flow fixed point and the stop-and-go wave are stable: $\rho\in(\rho^*_\text{SG1},\rho^*_\text{FF1})$ and $\rho\in(\rho^*_\text{FF2},\rho^*_\text{SG2})$. Similar bi-stable regimes were first recognized by \cite{gasser_bifurcation_2004}, \cite{orosz_subcritical_2006} and \cite{orosz_exciting_2009} in optimal velocity models using an ODE approach. The existence of bi-stability regimes implies that the linear stability (or string stability) does not always prevent the formation of stop-and-go waves.

\subsection{\bf The linearized map at the free-flow fixed point}\label{linearizedMap}
To investigate the linear stability of the map quantitatively, we linearize the map at the free-flow fixed point. Let's introduce small perturbations, $\eta_{i,t}\equiv \big(\phi_{i,t},\psi_{i,t},\theta_{i,t}\big)^\top$, around the free-flow fixed point as follows: $ \forall i\in\{1,\cdots,N\}$,
\begin{equation}
	\begin{cases}
		x_{i,t}&=x_{i,t}^* +\phi_{i,t}\,\,\,\pmod{C}\, ,\\
		v_{i,t}&=v_{i,t}^*+\psi_{i,t}\, , \\
		a_{i,t}&=a_{i,t}^*+\theta_{i,t}\, .
	\end{cases}\label{linearization}
\end{equation}
 Let's further define the slope vector of the control input with respect to the driver-observed state evaluated at the free-flow fixed point: $ \forall i\in\{1,\cdots,N\}$ and $l\in\{0,1\}$,
$$
	\beta_l=\big(\beta_l^x,\beta_l^v,\beta_l^a\big)
	\begin{array}[t]{l}
		\equiv\displaystyle\Big(\frac{\partial \bar{u}_{i,t}}{\partial x_{i+l,t}},
		\frac{\partial \bar{u}_{i,t}}{\partial v_{i+l,t}},
		\frac{\partial \bar{u}_{i,t}}{\partial a_{i+l,t}}\Big)(s^*_{i,t})\, .
	\end{array}
$$
Since all agents are identical, these $\beta$'s are independent of $i$ and $t$ at the fixed point. We also note that $\beta_0^x=-\beta_1^x$, because the utility only depends on the difference $x_{i+1,t}-x_{i,t}$. One can explicitly verify that the sign of each component of $\beta$ is consistent with the requirements for the acceleration function put forward by Treiber and Kesting \cite{treiber_traffic_2013} (see section 11.1): decreasing with own velocity 
($\beta_0^v<0$), increasing with headway ($\beta_1^x\ge 0$), and increasing with the velocity of the leading vehicle ($\beta_1^v\ge 0$).
Inserting these definitions into the full state evolution in Eq.(\ref{map_evolution}) and only keeping terms linear in perturbations we obtain the following linearized map: $ \forall i\in\{1,\cdots,N\}$,
\begin{equation}
	\begin{cases}
		\phi_{i,t+1}&=\phi_{i,t}+\psi_{i,t}\,\Delta t\,\,\,\pmod{C}\, ,\\
		\psi_{i,t+1}&=\psi_{i,t}+\theta_{i,t}\,\Delta t\, ,\\
		\theta_{i,t+1}&=\gamma\,\theta_{i,t}
		+\sum_{l=0,1}\beta_l\big(\eta_{i+l,t}-\gamma\eta_{i+l,t-1}\big) .
	\end{cases}
	\label{linearized_map}
\end{equation}

\subsection{\bf $z$-spectrum of the linearized map}
It is convenient to first perform a discrete Fourier transform on the $i$-index of the perturbative vectors $(\phi_{i,t},\psi_{i,t},\theta_{i,t})$, (i.e. $\phi_{i,t}=\frac{1}{N}\sum_{k=0}^{N-1}\tilde{\phi}_{k,t}\exp[\imath 2\pi k\,i/N]$, and so on), yielding the following decoupled linear map in frequency space:
\begin{equation}
	\begin{cases}
		\tilde{\phi}_{k,t+1}&=\tilde{\phi}_{k,t}+\tilde{\psi}_{k,t}\,\Delta t\, ,\\
		\tilde{\psi}_{k,t+1}&=\tilde{\psi}_{k,t}+\tilde{\theta}_{k,t}\,\Delta t\, ,\\
		\tilde{\theta}_{k,t+1}&=\gamma\,\tilde{\theta}_{k,t}
		+(\beta_0+\alpha_k\beta_1)\big(\tilde{\eta}_{k,t}-\gamma\tilde{\eta}_{k,t-1}\big)\, ,
	\end{cases}
	\label{map_fourier}
\end{equation}
where $\alpha_k=\exp(\imath\frac{2\pi k}{N})$, $k\in\{0,1,\cdots,N-1\}$, $\imath=\sqrt{-1}$, and
$\tilde{\eta}_{k,t}\equiv \big(\tilde{\phi}_{k,t},\tilde{\psi}_{k,t},\tilde{\theta}_{k,t}\big)^\top$.
Eq.(\ref{map_fourier}) is a second-order linear difference equation in $t$, hence it cannot be solved with the regular eigen-analysis. To investigate the stability of this linearized map with delay we make a unilateral $z$-transform in the $t$-index,   
(i.e. $\hat{\phi}_k(z)=\sum_{t=0}^\infty\tilde{\phi}_{k,t}\,z^{-t}$, and so on), on Eq.(\ref{map_fourier}).  After some algebra, we arrive at the following: $ \forall k\in\{0,1,\cdots,N-1\}$,
\begin{equation}
	P_k(z)\big(\hat{\phi}_k(z),\hat{\psi}_k(z),\hat{\theta}_k(z)\big)^\top=Q_k(z)\, .
\end{equation}
Note that the roots of the determinant of matrix $P_k(z)$ are the poles in the $z$-transformed state vector, and vector $Q_k(z)$ embodies the initial conditions of the $z$-transform. They can written as 
(with $B_k^x\equiv (\beta_0^x+\alpha_k\beta_1^x)$, and so on),
\begin{equation}
	P_k(z)=\\
	\begin{pmatrix}
		1-z & \Delta t & 0 \\ 0 & 1-z & \Delta t \\
		 (z-\gamma) B_k^x & (z-\gamma)B_k^v & (z-\gamma)(B_k^a -z)
	\end{pmatrix},
	\label{P_matrix}
\end{equation}
\begin{equation}
	Q_k(z)=\begin{pmatrix}
		(1-z)\tilde{\phi}_{k,0}-\tilde{\phi}_{k,1}+\Delta t\,\tilde{\psi}_{k,0} \\ 
		(1-z)\tilde{\psi}_{k,0}-\tilde{\psi}_{k,1}+\Delta t\,\tilde{\theta}_{k,0} \\ 
		(\gamma-z)z\,\tilde{\theta}_{k,0}-z\,\tilde{\theta}_{k,1}+
		z(\beta_0+\alpha_k\beta_1)\tilde{\eta}_{k,0}\big)
	\end{pmatrix} .
\end{equation}
Note also that $\det P_k(z)=0$ is analogous to the characteristic equation in the regular eigen-analysis, with the $z$-root playing the role of eigenvalue. 

Finally, the stability of the map is determined by the requirement that all roots of $\det P_k(z)=0$ lie within the unit circle (to ensure that the $z$-transform has no singularity in the domain $|z|>1$ where the power series of $z^{-1}$ can be convergent) \cite{lee_structure_2011}: $ \forall k\in\{0,1,\cdots,N-1\}$,
\begin{equation}
	(\gamma-z)\Big[(1-z)\Big((1-z)(z-B_k^a)+\Delta t\,B_k^v \Big)-(\Delta t)^2\,B_k^x\Big]=0.
	\label{map_roots}
\end{equation}
The above equation embodies $N$ independent equations (one for each Fourier mode).
First, we notice that Eq.(\ref{map_roots}) has $4N$ roots, in contrast with the situation with no delay (equivalent to the eigen-analysis) with only 3N roots. This is consistent with the fact that our map is truly $f: \mathbb{R}^{4N}\mapsto\mathbb{R}^{4N}$. Second, Eq.(\ref{map_roots}) has an $N$-fold degenerate root: $z=\gamma$, corresponding to an independent AR(1) process for each $z$-transformed Fourier mode $\hat{\theta}_{k}(z)$. Third, because $s_{i,t}$ enters the effective utility always in the combinations $\hat{x}_{l,1}=x_{l,t}+v_{l,t}\Delta t$ and $\hat{v}_{l,1}=v_{l,t}+a_{l,t}\Delta t$ for $l=i\text{ or }j$, it is easy to verify: $\beta_l^v-\beta_l^a/{\Delta t}=\Delta t\,\beta_l^x$. This relationship can be used to check that $z=0$ is also an $N$-fold degenerate root. Fourth,  there is a unit root ($z=1$) trivially when $k=0$, as a consequence of the translation invariance of the original dynamical system (or $B_{k=0}^x=0$).\footnote{If a root happened to pass through 1 when the externally influenceable parameters, such as $\rho$ and $v^*$, were dialed, then this would have signaled a fold bifurcation.} Because these three types of roots are independent of the externally influenceable parameters, $\rho$ and $v^*$, they are not related to the bifurcation phenomena we are trying to study. Hence, we will concentrate only on the remaining $(2N-1)$ non-trivial roots in our subsequent analysis. Fifth, since all components of $\beta$ are real, and $\alpha_{N-k}$ and $\alpha_k$ are conjugate to each other, it is easy to see that the roots have reflection symmetry with respect to the real axis, i.e. $z_{N-k}$ and $z_k$ are also conjugate to each other.  

\subsection{\bf Neimark-Sacker bifurcations}
While the linear stability of the original map is determined by the magnitude of roots of the linearized map, the nature of the bifurcation depends on how the roots cross the unit circle. When the vehicle density is low, all non-trivial roots lie inside the unit circle, indicating linear stability. The bifurcation arises when one or more roots touch the unit circle as the vehicle density increases. Further increasing the vehicle density will lead to the loss of linear stability, as some of the roots lie outside of the unit circle. 

\begin{figure}[!h]
	\centering
	\begin{tabular}{lr}
		\includegraphics[width=.46\textwidth]{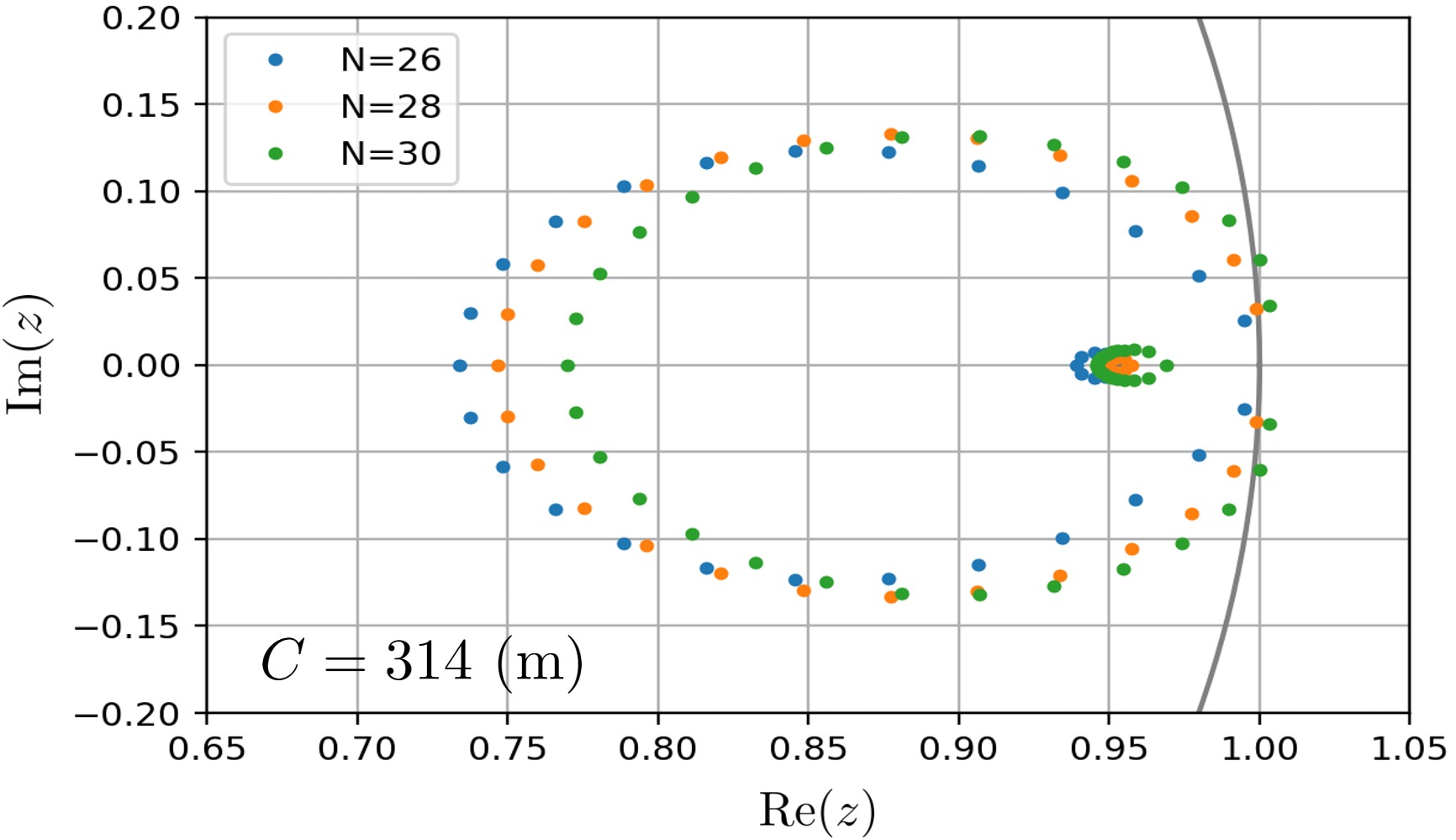}
	\end{tabular}
	\caption{Distribution of $(2N-1)$ non-trivial roots of $\det A_k(z)=0$ for the linearized map at the free-flow fixed point as a function of number of vehicles $N$,  with $C=314$ (m), corresponding to vehicle density near $\rho^*_\text{FF1}$. }
	\label{FF_roots_N}
\end{figure}

\begin{figure}[!h]
	\centering
	\begin{tabular}{lr}
		\includegraphics[width=.46\textwidth]{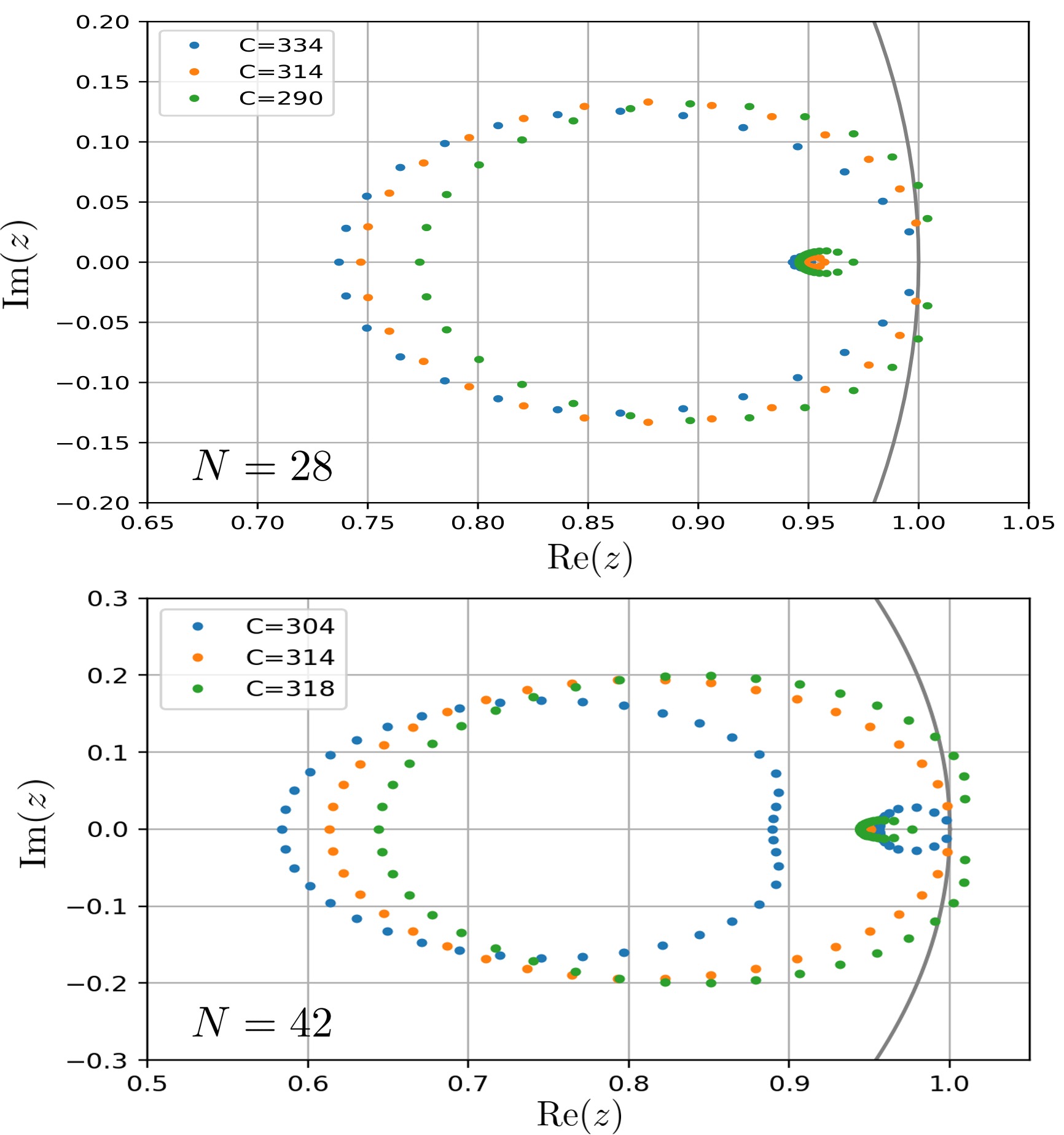}
	\end{tabular}
	\caption{Distribution of $(2N-1)$ non-trivial roots of $\det A_k(z)=0$ as a function of vehicle density $\rho$: (a) $N=28$ and three values of $C$, corresponding to vehicle density near $\rho^*_\text{FF1}$. The roots flow outwards from the unit circle as $\rho$ increases. (b) $N=42$ and three values of $C$, corresponding to vehicle density near $\rho^*_\text{FF2}$. The roots flow inwards to the unit circle as $\rho$ increases.}
	\label{FF_roots_C}
\end{figure}

In Fig.\ref{FF_roots_N} we plot the distribution of $(2N-1)$ non-trivial $z$-roots for the varied numbers of vehicles, $N$ = 26, 28, or 30, with the fixed circumference $C=314$ (m) corresponding to vehicle density near $\rho^*_\text{FF1}$.  When $N = 26$,  all the roots are within the unit circle, indicating that the free-flow fixed point is linearly stable. When $N=28$ a pair of non-trivial complex roots that are conjugate to each other touches the unit circle. This is where Neimark-Sacker bifurcation occurs. The bifurcation is called Neimark-Sacker because the aforementioned reflection symmetry guarantees that a pair of complex conjugate roots always crosses the unit circle simultaneously, analogous to the Hopf bifurcation in the ODEs of the models considered by \cite{gasser_bifurcation_2004, orosz_subcritical_2006, orosz_exciting_2009}. When $N=30$ there are two pairs of complex conjugate roots that lie outside the unit circle and hence the free-flow fixed point of the map becomes linearly unstable.

In Fig.\ref{FF_roots_C} we plot the distribution of $(2N-1)$ non-trivial $z$-roots for various values of circumferences,  with a fixed number of vehicles.  Because it can be varied continuously,  $C$ should be tuned to locate the critical density more precisely.  As the circumference decreases (or $\rho$ increases) the roots flow outwards from the unit circle when $N=28$ (a),  corresponding to vehicle density near $\rho^*_\text{FF1}$; and the opposite is true when $N=42$ (b),  corresponding to vehicle density near $\rho^*_\text{FF2}$. The first critical vehicle density is estimated to be at around $\rho^*_\text{FF1}\approx 0.090$ (1/m) (above which the system becomes linearly unstable), and the second critical vehicle density at $\rho^*_\text{FF2}\approx 0.134$ (1/m) (above which the system restores its linear stability). To further elucidate the finer details of the bifurcation, such as whether the bifurcation is subcritical or supercritical, requires analyzing higher order effects using the so-called normal form \cite{kuznetsov_elements_2004}. We will leave that for future study.

Another useful fact to notice is that the lowest few Fourier modes (with low-$k$ values and their conjugates) are the ones that are responsible for the Neimark-Sacker bifurcation.

\section{\bf Bifurcation diagram: {\it simulation study}}\label{bistable_regime}
As pointed out by \cite{gasser_bifurcation_2004, orosz_subcritical_2006, orosz_exciting_2009}, stop-and-go phantom waves develop even before the Hopf bifurcation is reached when vehicle density gradually increases. This is because there is another branch of asymptotically stable limit solutions with high oscillation magnitude $A\equiv\text{mean}_t\big(\max_i v_{i,t}-\min_i v_{i,t}\big)$. As shown by \cite{orosz_exciting_2009}, the nature of the associated bifurcation of this branch is a fold, which is different from the Hopf bifurcation associated with the free-flow fixed point. Following a method deployed in \cite{orosz_exciting_2009}, we use simulation to semi-quantitatively investigate where the new branch is located in the bifurcation diagram. This approach is feasible because the quasi-period orbit corresponding to the phantom wave is asymptotically stable.  To further characterize the asymptotic solution, another order parameter,  average velocity $V\equiv\displaystyle{\text{mean}_{i,t}(v_{i,t})}$, will also be very useful.

\begin{figure}[!h]
	\centering
	\begin{tabular}{lr}
		\includegraphics[width=0.44\textwidth]{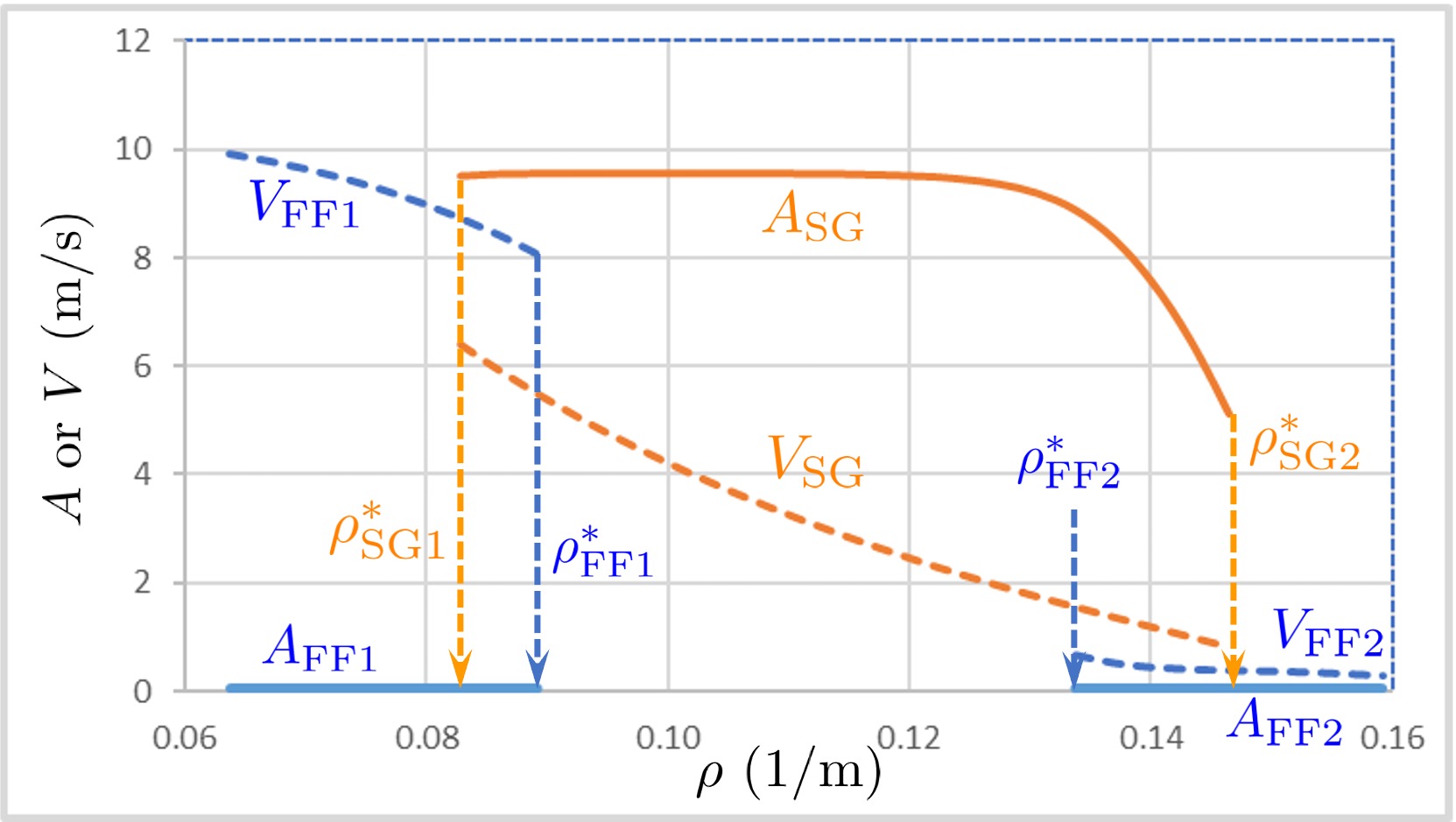}
	\end{tabular}
	\caption{Bifurcation diagram (varying $N$ and $C$ at around $N=28$ or $N=42$ and $C=314$ (m) at fixed $v^*=10.49$ (m/s)): oscillation amplitude $A$ (full lines) and average velocity $V$ (dashed lines) vs. vehicle density $\rho$. The asymptotically stable solution of stop-and-go waves is represented in orange, while the asymptotically stable solution of free flow is represented in blue. Bi-stable regimes exist for $\rho\in(\rho^*_\text{SG1},\rho^*_\text{FF1})$ and $\rho\in(\rho^*_\text{FF2},\rho^*_\text{SG2})$.}
	\label{bifurcation_plot}
\end{figure}
We simulate the discrete time dynamical system with $v^*=10.49$ (m/s) by following the same procedure outlined in Section \ref{limit_solutions}. The resulting bifurcation diagram is shown in Fig.\ref{bifurcation_plot}. The estimated critical points for the onset of the asymptotically stable solution corresponding to stop-and-go waves are at $\rho^*_\text{SG1}\approx 0.082$ (1/m) and $\rho^*_\text{SG2}\approx 0.146$ (1/m). Orosz et al \cite{orosz_subcritical_2006, orosz_exciting_2009} had showed, using the numerical continuation technique, that the stop-and-go solution is a large-amplitude periodic oscillation that becomes linearly stable at a fold bifurcation point $\rho^*_\text{SG1}$, with the lowest Fourier mode $k=1$ in a specific optimal velocity car-following model. Unfortunately, since we are relying on a simulation technique it is not possible to pin down the nature of the points $\rho^*_\text{SG1}$ and $\rho^*_\text{SG2}$ in our model. Even though the oscillation amplitudes appear to be constant until $\rho$ is close to $\rho^*_\text{SG2}$, the average velocities monotonically decrease as functions of the vehicle density. 

We note from Fig.\ref{bifurcation_plot} that the two bi-stable regimes are qualitatively different. On the lower vehicle density side $\rho\in(\rho^*_\text{SG1},\rho^*_\text{FF1})$, the free-flow solution is superior in both average speed (higher is better) and speed variation (lower is better) relative to the stop-and-go wave solution, i.e., $V_\text{FF1} > V_\text{SG}$ and $A_\text{FF1} < A_\text{SG}$. In the higher vehicle density side $\rho\in(\rho^*_\text{FF2},\rho^*_\text{SG2})$, the free-flow solution is smoother (with zero speed variation) but less efficient (lower average speed) relative to the stop-and-go wave solution, i.e.,  $A_\text{FF2} < A_\text{SG}$ but $V_\text{FF2} < V_\text{SG}$. From the point of view of traffic management, we are mostly interested in the first bi-stable regime where the free-flow fixed point solution is a Pareto improvement against the stop-and-go wave solution. The second bi-stable regime is so crowded that the average headway is less than the length of single car, thus there is little room left for traffic control to maneuver.  In addition,  it is possible that these limit solutions in the higher density regime might not be stable against perturbations in driver heterogeneity and state evolution noise.

It is somewhat surprising that the bifurcation diagram for the {\it adaptiveSeek} map is so simple, relative to those found in much simpler maps, such as the delayed logistic map and the Henon map. Though we made some effort in experimenting with the initial conditions in various simulations, we did not find any other types of  asymptotically stable solutions that are qualitatively different from the free-flow fixed point and the stop-and-go quasi-periodic wave. In particular, we did not observe any candidate solutions that behave chaotically. Of course, this does not mean that they do not exist. It only means that if there are chaotic solutions, the basins of attraction are relatively small. Subsequently, the effect of other possible solutions is minor on practical traffic management.  

\section{\bf Traffic flow phase diagram: {\it 2D bifurcation analysis}}\label{2D_bifurcation}
So far we have been concentrating on studying the nature of traffic flow as a function of vehicle density, while keeping all the model parameters fixed at the values calibrated using vehicle trajectory data from the Sugiyama experiment. The bifurcation diagram in Fig.\ref{bifurcation_plot} is only one-dimensional in vehicle density. We now shift our attention to how we might tame the phantom wave
with some external control. More specifically, we want to investigate the mechanism found in \cite{shen_taming_2021} where the ideal speed can be dialed down under a smart speed advisory scheme. Our explicit purposes are 1) to see how the bifurcation points are altered when the speed advisory restricts the ideal speed to be sufficiently low, and 2) to sketch out the traffic flow phase diagram in the $(\rho,v^*)$-plane for an optimal solution for taming the phantom wave via variable speed advisory. 

\begin{figure}[!h]
	\centering
	\includegraphics[width=.45\textwidth]{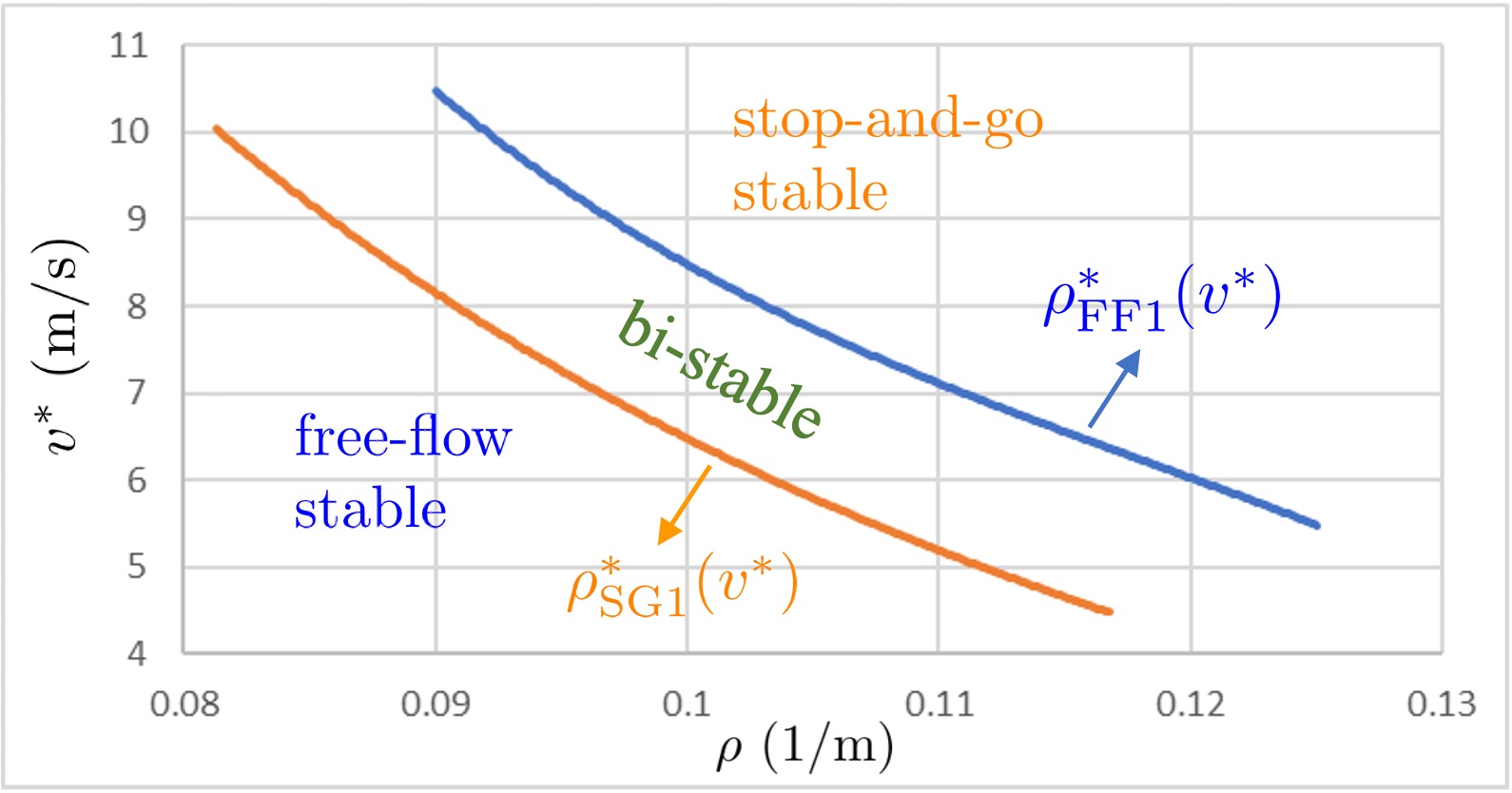}
	\caption{Traffic flow phase diagram in the $(\rho,v^*)$-plane: $\rho^*_\text{SG1}(v^*)$ and $\rho^*_\text{FF1}(v^*)$ as functions of the restricted ideal speed. The region below the orange curve is the free-flow phase. The region above the blue curve is the stop-and-go phase. The region between the orange and blue curves is bi-stable where both free-flow and stop-and-go solutions can happen.}
	\label{phase_diagram}
\end{figure}

In Fig.\ref{phase_diagram} we show numerical results of the two bifurcation points: $\rho^*_\text{SG1}(v^*)$ and $\rho^*_\text{FF1}(v^*)$, as functions of the ideal speed lower than the calibrated value of $v^*=10.49$ (m/s).  This is a sort of two-dimensional bifurcation diagram. The blue curve $\rho^*_\text{FF1}(v^*)$ and the orange curve $\rho^*_\text{SG1}(v^*)$ partition the $(\rho,v^*)$-plane into three regions. The free-flow phase lies below $\rho^*_\text{SG1}(v^*)$, the stop-and-go phase above $\rho^*_\text{FF1}(v^*)$, and the bi-stability phase in between. It is much easier to get the blue curve of $\rho^*_\text{FF1}(v^*)$, as we only need to find the first pair of $z$-roots crossing the unit circle. Because our method relies on detecting stop-and-go like limit solutions using very long simulations, the orange curve is much harder to pin down.  Therefore,  the orange curve in the figure should be viewed as a upper bound on the true $\rho^*_\text{SG1}(v^*)$.  

For the purpose of taming the stop-and-go phantom wave, the recommended speed should be chosen just below the orange curve in Fig.\ref{phase_diagram} for each given vehicle density, so that the dynamical system stays entirely in a free-flow regime.  Equivalently,  the solution of $\rho=\rho^*_\text{SG1}(v^*_c)$ implicitly defines  a critical ideal speed curve $v^*_c(\rho)$ below which the traffic system is kept in a free-flow phase at $\rho$.
Interestingly,  the orange curve here is almost the same as the optimal ideal speed curve of Fig.7 in \cite{shen_taming_2021} (colored purple and labeled by N-CAV there), in which driver heterogeneity and state evolution noise were explicitly included. This close agreement indicates that the optimal ideal speed advisory curve, found by optimizing an aggregate objective function in the computational mechanism design of the traffic management authority, is essentially the ideal speed curve $v^*_c(\rho)$ implied by $\rho=\rho^*_\text{SG1}(v^*_c)$.

\begin{figure}[!h]
	\centering
	\includegraphics[width=.45\textwidth]{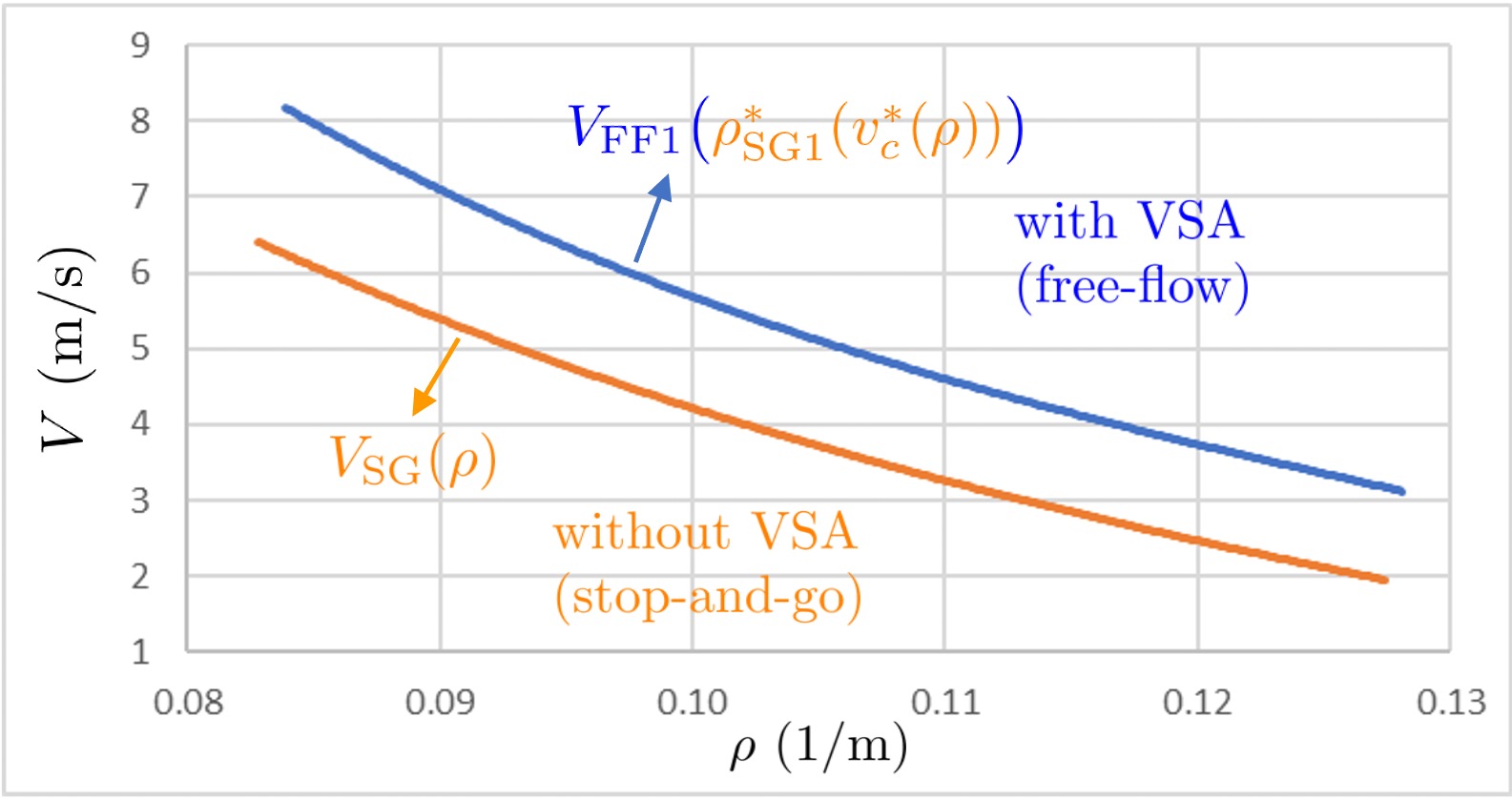}
	\caption{Average velocity of the traffic flow with VSA (blue), i.e.  $V_\text{FF1}\big(\rho^*_\text{SG1}(v^*_c(\rho))\big)$,  and without VSA (orange) (same as $V_\text{SG}(\rho)$ in Fig.\ref{bifurcation_plot}).  Also, with a VSA the oscillation amplitude vanishes,  in contrast with that without a VSA (same as $A_\text{SG}$ in Fig.\ref{bifurcation_plot}).}
	\label{avg_speed}
\end{figure}

In Fig.\ref{avg_speed} we show, again using simulation, the average velocity of the traffic flow with and without VSA, demonstrating that imposing a vehicle-density dependent speed advisory substantially improves traffic flow, both in efficiency (higher average velocity) and smoothness (lower speed variation). The magnitude of the improvement is quantitatively similar with Fig.8 of \cite{shen_taming_2021} (the cases with N-CAV),  where the same behavior model is used but with explicit driver heterogeneity and state evolution noise.

\section{\bf Summary and Conclusions}
We have performed a bifurcation analysis on a map of a discrete-time dynamical system for car-following traffic on a circular road.  The map corresponds to a human driving behavior model, whose driving policy is derived from maximizing a judicially chosen utility function.  After its parameters are systematically calibrated using vehicle trajectory data from the Sugiyama experiment, the model is shown to closely reproduce the observed collective traffic patterns from the more general Tadaki experiment.  When only vehicle density is varied, two asymptotically stable limit solutions were found, one fixed point representing free-flow traffic, and another quasi-period orbit representing stop-and-go waves.  Despite the apparent difference in formulation, we found very similar bifurcation diagrams to those found with driving policy-based models using ODEs, 
including the nature of the bifurcation under which the system loses its linear stability, as well as the existence of bi-stable regimes. We further extended our bifurcation analysis into the joint space of vehicle density and ideal speed, with the latter being enforced by a VSA. The traffic flow phase diagram in the $(\rho,v^*)$-plane was then estimated using simulations.  We demonstrate numerically that the bifurcation curve $\rho^*_\text{SG1}(v^*)$ is quantitatively very similar to the optimal traffic control curve for taming stop-and-go waves using a computational mechanism design approach \cite{shen_taming_2021}. Therefore, our work presented here provides a solid theoretical foundation for the simple mechanism proposed in \cite{shen_taming_2021}.

A few mathematical details can be investigated further.  It may be of interest to know the precise nature of the Neimark-Sacker bifurcation, such as whether it is sub-critical or supercritical.  Are there any other asymptotically stable solutions beyond the free-flow fixed point and stop-and-go wave? Perhaps more interestingly,  it is natural to ask whether emerging connectivity and edge computing, and resulting vehicle-to-vehicle coordination, such as along the line of the CIRCLES project \cite{bayen_circles_nodate},  can provide a better way to tame stop-and-go waves.  One such possibility is to iterate {\it adaptiveSeek} so that vehicles can explicitly negotiate, provided high-frequency communication can be established (either among vehicles or with a central authority). The iterated {\it adaptiveSeek} is a good approximation to a Nash equilibrium solution among all agents.  If so, this would be equivalent to pushing the $\rho^*_\text{SG1}(v^*)$-curve up and to the right, making the free-flow stability region larger in the $(\rho,v^*)$-plane.

\bibliographystyle{IEEEtran}
\bibliography{String_instability}

\end{document}